\documentclass[aps,prd,reprint,superscriptaddress,nofootinbib,floatfix]{revtex4-2}

\usepackage{amsmath,amssymb,bm}
\usepackage{graphicx}
\usepackage{booktabs}
\usepackage{float}
\usepackage{xcolor}
\usepackage[resetlabels]{multibib}
\usepackage[hidelinks]{hyperref}

\newcommand{\Xipp}{\Xi_{cc}^{++}}
\newcommand{\Xip}{\Xi_{cc}^{+}}
\newcommand{\Tu}{T_u}
\newcommand{\Td}{T_d}
\newcommand{\MeV}{\mathrm{MeV}}
\newcommand{\GeV}{\mathrm{GeV}}

\newcommand{\Br}{\mathcal B}
\newcommand{\Lag}{\mathcal L}
\newcommand{\order}{\mathcal O}

\begin{document}

\title{$\Xi_{cc}^{++}-\Xi_{cc}^{+}$ Transitions as a Two-Charm-Selective Portal to Ultra-Low-$Q$ Charged Currents}

\author{Yong Du}\email{yongdu5@impcas.ac.cn} 
\affiliation{Institute of Modern Physics, Chinese Academy of Sciences, Lanzhou 730000, China} 
\affiliation{School of Nuclear Science and Technology, University of Chinese Academy of Sciences,\\
	19A Yuquan Road, Beijing 100049, China}  

\begin{abstract}
The recent LHCb observation of $\Xi_{cc}^{+}$ opens the ultra-low-$Q$ transition $\Xi_{cc}^{++}\to\Xi_{cc}^{+}$ as an experimentally motivated null test of charged-current new physics. Our two- and three-body analyses show sensitivity to effective baryon-level couplings of $\order(10^{-6}-10^{-7})$ for MeV-scale recoil momenta. We establish a practical no-go result for a universal light charged scalar $\phi^+$: with a first-generation $\phi^+\bar u d$ coupling, existing electroweak-precision and beta-decay constraints are parametrically stronger than the projected LHCb sensitivity. We then identify a two-charm-selective charged-current portal whose leading operator has a nonzero matrix element in doubly charmed baryons but vanishes at leading order in pions, nucleons, nuclei, and singly charmed mesons. In this class of models, $\Xi_{cc}^{++}\to\Xi_{cc}^{+}$ transitions can provide the leading direct probe of the portal interaction.
\end{abstract}

\maketitle

\section{Introduction}
The doubly charmed isodoublet
\begin{equation}
  \Tu\equiv\Xipp(ccu),\qquad \Td\equiv\Xip(ccd)
\end{equation}
provides a heavy-hadron analogue of beta decay in which the compact $cc$ pair is a spectator. The LHCb Collaboration~\cite{LHCb:2026pxn} recently observed $\Xip$ in Run-3 data and measured
\begin{equation}
  M(\Xip)-M(\Xipp)= -1.77\pm0.84\pm0.15{}^{+1.90}_{-1.30}\,\MeV.
  \label{eq:LHCbDelta}
\end{equation}
Thus, at the baseline lifetime correction used by LHCb, we define
\begin{equation}
  Q_\Xi\equiv M(\Xipp)-M(\Xip)=1.77\,\MeV.
  \label{eq:Qdef}
\end{equation}
The final asymmetric uncertainty in Eq.\,\eqref{eq:LHCbDelta} arises from the assumed $\Xip$ lifetime, $\tau(\Xip)$, which enters the correction for a selection-induced mass bias. LHCb adopts $\tau(\Xip)=45\,{\rm fs}$ as its baseline, motivated by the heavy-quark-expansion (HQE) estimate of Ref.~\cite{ChengShi:2018doublyLifetime}, and scans the range $15$--$160\,{\rm fs}$. More recent HQE calculations~\cite{Cheng:2026mlv,Dulibic:2026atz} provide updated lifetime predictions that can be used to sharpen the corresponding uncertainty in $Q_\Xi$.

Independent phenomenological calculations also favor a positive, MeV-scale mass difference\,\cite{Hwang:2008dj,Brodsky:2011zs,WeiChenGuo:2015xqa,KarlinerRosner:2017xqf}, and a fully dynamical QCD+QED lattice calculation by the BMW Collaboration gives~\cite{Borsanyi:2014jba} $Q_\Xi=2.16(11)(17)\,\MeV$. We therefore use the LHCb central value, $Q_\Xi=1.77\,\MeV$, as our benchmark, while retaining the explicit $Q_\Xi$ dependence wherever it is phenomenologically relevant.

Because $Q_\Xi\ll m_\pi$, ordinary hadronic transitions are kinematically forbidden.  The only open Standard Model (SM) channel, $\Xipp\to\Xip e^+\nu_e$, is strongly suppressed by $Q_\Xi^4G_F^2$ with $G_F$ the Fermi constant and is expected to have a branching fraction of $\order(10^{-21})$. We therefore consider the net-charge-one final-state systems
\begin{equation}
  \Xipp\to\Xip {\cal P},
  \label{eq:signalmain}
\end{equation}
where ${\cal P}$ denotes either a reconstructed charged recoil $X^+$ or a semileptonic system $e^+N$ with $N$ a light neutral fermion. Observation of the two-body mode, or of a three-body rate or kinematic distribution inconsistent with the tiny SM beta-decay contribution, would constitute a null test of the SM.

\section{Model-independent reach}
\begin{figure*}[!thb]
  \centering
  \includegraphics[width=0.48\linewidth]{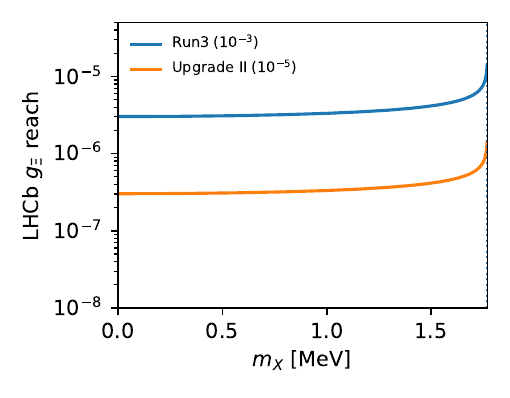}
  \includegraphics[width=0.48\linewidth]{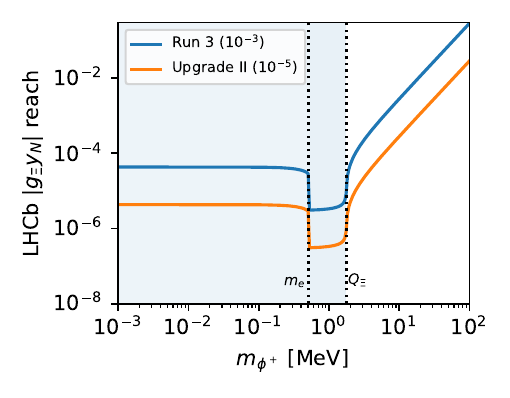}
  \caption{Left: projected LHCb sensitivity to the two-body coupling $g_\Xi$ for $Q_\Xi=1.77\,\MeV$. Right: projected sensitivity to the three-body final state mediated by a charged scalar $\phi^+$. We set $m_N=0$ and assume $\phi^+\to e^+N$ saturates the $\phi^+$ width in the on-shell region between the vertical dashed lines. The two benchmark sensitivities are defined in the main text.}
  \label{fig:twobody}
\end{figure*}

At the baryon level, we parameterize emission of a spin-0 charged state by
\begin{equation}
  \Lag_X=X^+\,\overline{\Xip}(g_\Xi+i g_P\gamma_5)\Xipp+{\rm h.c.},
  \label{eq:baryonmain}
\end{equation}
where $g_\Xi$ and $g_P$ are scalar and pseudoscalar baryonic couplings, respectively. Near threshold,
\begin{equation}
 \Gamma(\Xipp\to\Xip X^+)\simeq \frac{|g_\Xi|^2}{2\pi} \, \sqrt{Q_\Xi^2-m_X^2}.
 \label{eq:widthmain}
\end{equation}
Using $\tau(\Xipp)=0.256^{+0.024}_{-0.022}\pm0.014\,{\rm ps}$~\cite{LHCb:2018xiccLifetime}, one finds
\begin{equation}
  \Br_X\simeq6.2\times10^7 |g_\Xi|^2\left(\frac{\sqrt{Q_\Xi^2-m_X^2}}{1\,\MeV}\right),
  \label{eq:BRmain}
\end{equation}
where $\Br_X\equiv\Br(\Xipp\to\Xip X^+)$ is the branching fraction. The left panel of Fig.\,\ref{fig:twobody} shows the projected LHCb sensitivity for the two benchmark reaches defined in the next section. The Run-3 benchmark probes $g_\Xi$ at the $\order(10^{-6})$ level in this model-independent two-body parametrization, with an improvement of approximately one order of magnitude at Upgrade II.

The same ultra-low-$Q$ phase-space window also permits the three-body semileptonic decay
\begin{equation}
 \Xipp\to\Xip e^+N.
 \label{eq:eNmain}
\end{equation}
Introducing a charged scalar mediator $\phi^+$ provides a unified description of the heavy-mediator, resonant, and light-mediator regimes:
\begin{equation}
  \Lag_{\phi eN}=g_\Xi \phi^+\overline{\Xip}\Xipp+y_N\phi^+\bar N P_R e+{\rm h.c.},
  \label{eq:phiPropMain}
\end{equation}
where $g_\Xi$ is the same effective $\Xipp$--$\Xip$ scalar coupling as in Eq.\,\eqref{eq:baryonmain}. The differential rate takes the Breit--Wigner form
\begin{equation}
 \frac{d\Gamma_{eN}^{(\phi)}}{ds}=\frac{1}{\pi}
 \frac{\sqrt{s}\,\Gamma_{\Xi\to\Xi\phi^*(s)}\,\Gamma_{\phi^*(s)\to eN}}
 {(s-m_\phi^2)^2+m_\phi^2\Gamma_\phi^2},
 \label{eq:propMain}
\end{equation}
where $s\equiv(p_e+p_N)^2$. For $m_N=0$, the off-shell partial widths are
\begin{align}
 \Gamma_{\Xi\to\Xi\phi^*(s)}&\simeq \frac{|g_\Xi|^2}{2\pi}\sqrt{Q_\Xi^2-s},
 \\
 \Gamma_{\phi^*(s)\to eN}&=\frac{|y_N|^2\sqrt{s}}{16\pi}\left(1-\frac{m_e^2}{s}\right)^2.
 \label{eqS:partialWidthsPhi}
\end{align}

The right panel of Fig.\,\ref{fig:twobody} shows the corresponding LHCb sensitivity. For $m_\phi<m_e$, the Run-3 benchmark probes $|g_\Xi y_N|$ at the $10^{-5}$ level, with an improvement of approximately one order of magnitude at Upgrade II. In the on-shell region, $m_e+m_N<m_\phi<Q_\Xi$, the narrow-width approximation factorizes the rate into $\Xipp\to\Xip\phi^+$ followed by $\phi^+\to e^+N$; the plotted reach assumes $\Br(\phi^+\to e^+N)=1$. For $m_\phi>Q_\Xi$, the mediator is off shell and the sensitivity degrades rapidly. The ultra-low-$Q$ $\Xipp\to\Xip$ transition is therefore most sensitive to light charged-current mediators.

\section{Experimental observables at LHCb}
If $X^+$ is reconstructed, the natural visible-recoil observable is the reconstructed mass difference
\begin{equation}
 \Delta m_X=m(\Xip X^+)-m(\Xip).
 \label{eq:deltamobs}
\end{equation}
For a correctly reconstructed two-body signal, this observable peaks at $M(\Xipp)-M(\Xip)=Q_\Xi$. The kinematically small quantities are instead the kinetic-energy release $Q_\Xi-m_X$ and the recoil momentum in the $\Xipp$ rest frame. Direct reconstruction of a singly charged MeV-scale state is, however, severely efficiency-suppressed at LHCb: even a $50\,\GeV$ $\Xipp$ boosts a MeV-scale recoil only to momenta of $\order(10\,\MeV)$, far below the $\sim1.5\,\GeV$ scale required for a charged particle to reach the downstream tracking stations~\cite{LHCb:2015DetectorPerformance}. A more promising strategy is therefore to search for a feed-down component in the reconstructed $\Xip\to\Lambda_c^+K^-\pi^+$ sample. Promptly produced $\Xip$ baryons and cascade $\Xip$ baryons from $\Xipp\to\Xip X^+$ populate the same $\Lambda_c^+K^-\pi^+$ mass peak, but differ in impact-parameter and decay-time-proxy distributions because the cascade component contains the additional upstream $\Xipp$ flight. A multicomponent template fit to the $\Xip$ mass and topology variables, normalized to the abundant $\Xipp$ control mode, can therefore constrain $R_X$ even when $X^+$ is not reconstructed. For $X^+\to e^+\nu$ or the decay in Eq.\,\eqref{eq:eNmain}, the visible invariant-mass spectrum is continuous because the neutral fermion is unobserved. If the parent flight direction can be inferred, a corrected-mass analysis can nevertheless provide an endpoint observable, as in standard LHCb semileptonic analyses~\cite{LHCb:2020PpbarMuNu}.

Accordingly, we define product-ratio observables normalized to established LHCb reconstruction modes:
\begin{equation}
  R_{\cal P}=\frac{\Br(\Xipp\to\Xip {\cal P})\Br(\Xip\to\Lambda_c^+K^-\pi^+)}{
  \Br(\Xipp\to\Lambda_c^+K^-\pi^+\pi^+)},
  \label{eq:RXmain}
\end{equation}
with ${\cal P}=\{X^+,\,e^+N\}$. The absolute branching fraction is
\begin{align}
\begin{split}
 &\, \Br(\Xipp\to\Xip {\cal P})=\rho_B R_{\cal P},\\
 &\, \rho_B \equiv
  {\Br(\Xipp\to\Lambda_c^+K^-\pi^+\pi^+)\over
  \Br(\Xip\to\Lambda_c^+K^-\pi^+)} .
  \label{eq:rhoBmain}
\end{split}
\end{align}
All coupling-reach curves use the convention $\rho_B=1$ and for another value, the LHCb sensitivity reach rescales as $\rho_B^{1/2}$. Let $N_{++}$ denote the reconstructed yield in the normalization mode $\Xipp\to\Lambda_c^+K^-\pi^+\pi^+$ and $\epsilon_{\rm rel}$ the signal-to-normalization efficiency ratio. The expected signal yield is
\begin{equation}
  s_{\cal P}=N_{++}\,R_{\cal P}\,\epsilon_{\rm rel},\qquad R_{\cal P}^{95}\simeq \frac{3.0}{N_{++}\epsilon_{\rm rel}}
  \label{eq:Poissonmain}
\end{equation}
for a zero-background 95\% C.L. estimate. The publicly reported Run-3 normalization yield is $N_{++}=8712\pm160$ in $6.9\,{\rm fb}^{-1}$~\cite{Han:2026MoriondXicc}.

We estimate Eq.\,\eqref{eq:Poissonmain} with \textsc{GenXicc}-based\,\cite{Chang:2009va} fast-simulation using a $10^8$-event $\Xipp$ production sample and a two-dimensional topology-template fit separating prompt $\Xip$ production from cascade $\Xip$ feed-down\,\cite{Chang:2009va,LHCb:2017iph,LHCb:2026pxn}. For $Q_\Xi=1.77\,\MeV$, the benchmark two-body point $m_X=1.0\,\MeV$ gives the feed-down efficiency relative to the normalization mode as $\epsilon_{\rm rel}^{\rm fd}=0.571$, and
\begin{align}
  R_{X,{\rm Run\,3}}^{95} &={1.31\times10^{-3}},
  & R_{X,{\rm II}}^{95} &={3.72\times10^{-5}},
  \label{eq:fastsimR95MainX}
\end{align}
while a constant-matrix-element $e^+N$ benchmark with $m_N=0$ gives
\begin{align}
  R_{eN,{\rm Run\,3}}^{95} &={1.35\times10^{-3}},
  & R_{eN,{\rm II}}^{95} &={3.55\times10^{-5}}.
  \label{eq:fastsimR95MainEN}
\end{align}

Direct reconstruction of the MeV recoil or soft positron is strongly disfavored: the visible-recoil trackability fraction is $f_{\rm trk}(X^+)=2.99\times10^{-4}$ at the two-body benchmark, and the median laboratory-frame positron momentum in the $e^+N$ benchmark is only $9.4\,\MeV$. The Run-3 feed-down reach remains at the $10^{-3}$ level under topology and efficiency stress tests, whereas the Upgrade-II number is conditional on controlling signal-like feed-down backgrounds at the $10^{-4}$ level. We therefore use $R_{\cal P}^{95}=10^{-3}$ as a robust Run-3 target and $R_{\cal P}^{95}=10^{-5}$ as a conditional Upgrade-II benchmark; the numerical grids, template definitions, and stress tests are given in Appendix\,\ref{app:lhcbworkflow}.

\section{No-go result for universal light charged scalars}\label{sec:nogo}
The preceding discussion treats $X^+$ as a generic charged state. We now show that, if $X^+$ is identified with a light scalar $\phi^+$ carrying a universal first-generation charged-current coupling, LHCb cannot provide the leading constraint. We write
\begin{equation}
  \Lag_{ud\phi}=\phi^+\bar u(y_LP_L+y_RP_R)d+{\rm h.c.}\,,
  \label{eq:universalmain}
\end{equation}
as the simplest microscopic realization of Eq.\,\eqref{eq:signalmain}. First, an elementary charged scalar in a conventional electroweak representation with $m_\phi<M_Z/2$ contributes to the $Z$ width,
\begin{align}
 \Gamma(Z\to\phi^+\phi^-)\simeq
 \frac{G_FM_Z^3}{6\sqrt2\pi}q_Z^2
 \left(1-\frac{4m_\phi^2}{M_Z^2}\right)^{3/2},
 \label{eq:Zwidthmain}
\end{align}
with $q_Z\equiv T_3-Qs_W^2$. Here, $T_3$ and $Q$ are the weak isospin and electric charge of $\phi^+$ in units of the proton charge, and $s_W\equiv\sin\theta_W$. For conventional electroweak-singlet or -doublet assignments, Eq.\,\eqref{eq:Zwidthmain} gives an additional partial width of $\order(10\,\MeV)$, which is incompatible with the measured total $Z$ width~\cite{ALEPH:2005ab,PDG:2024}. Independently, a charged scalar below a few MeV is excluded in standard thermal cosmology~\cite{Escudero:2018mvt,Du:2021idh,Du:2023upj}.

Second, even if a nonstandard completion evades the electroweak and cosmological bounds, the same universal $\bar u d$ interaction induces zero-charm processes: $\pi^+\to\pi^0\phi^+$, $n\to p\phi^-$ when kinematically allowed, and exotic two-body decays of superallowed parent nuclei. We express these processes in terms of a common effective zero-charm coupling $g_{\rm ord}$,
\begin{equation}
  \Gamma_i^\phi=
  \frac{C_i|g_{\rm ord}|^2p_i}{2\pi},
  \qquad
  p_i\simeq\sqrt{Q_i^2-m_\phi^2},
  \label{eq:ordinaryWidth}
\end{equation}
where $C_i$ encodes the channel-dependent spin, isospin, and hadronic or nuclear matrix elements. We construct a joint profile likelihood from the PIBETA measurement of pion beta decay, the neutron lifetime, and the 15 best-measured superallowed $0^+\!\to0^+$ transitions in the Hardy--Towner survey~\cite{Pocanic:2003pf,PDG:2024,Hardy:2020qwl}. A common $V_{ud}$ parameter and inner radiative correction $\Delta_R^V$ are profiled across the three data sets, while the superallowed-transition likelihood retains the correlated $\delta_R'$ and $\delta_{NS}$ uncertainties. Details and numerical inputs are given in Appendix\,\ref{app:pionlikelihood}.

Fig.\,\ref{fig:ordinaryMain} shows the resulting mass-dependent constraint.  For the benchmark $C_i=\alpha_i=1$, the combined fit gives
\begin{align}
  g_{\rm ord}^{95}(m_\phi=0)=1.37\times10^{-13},
  \label{eq:ordinaryCombinedMain}
\end{align}
and the best fit is statistically consistent with the null hypothesis. Below the neutron threshold, the pion, neutron, and nuclear data contribute jointly; above threshold, the correlated superallowed-transition likelihood dominates. Comparing Eq.\,\eqref{eq:ordinaryCombinedMain} with the projected LHCb sensitivity, $g_\Xi\sim10^{-6}$--$10^{-7}$, shows that the zero-charm constraint lies approximately six to seven orders of magnitude below the coupling accessible in $\Xipp\to\Xip\phi^+$, provided $g_{\rm ord}\sim g_\Xi$ up to hadronic matching factors. The numerical result is conditional on the response coefficients $\alpha_i$ and matrix-element normalizations $C_i$, but the hierarchy is sufficiently large that LHCb cannot provide the leading constraint on a universal light scalar.

\begin{figure}[t]
  \centering
  \includegraphics[width=0.95\linewidth]{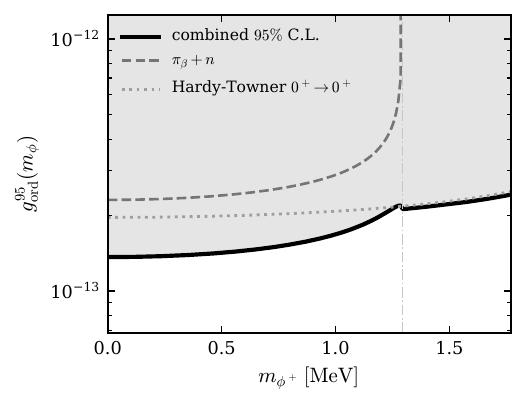}
  \caption{One-sided 95\% C.L. profile-likelihood limit on the universal zero-charm coupling. The black curve shows the combined pion, neutron, and correlated superallowed-transition result; the dashed and dotted curves show the pion-plus-neutron and nuclear sub-likelihoods, respectively. The shaded region is excluded for the benchmark normalization $C_i=\alpha_i=1$, where $\alpha_i$ parameterizes the response of the extracted superallowed observable to an unobserved exotic branch. The kink at $m_\phi=m_n-m_p$ marks the closure of the free-neutron channel.}
  \label{fig:ordinaryMain}
\end{figure}

\section{Two-charm-selective benchmark}
The no-go result can be avoided if the $u\to d$ current is proportional to a local two-charm structure. In a model-independent effective field theory (EFT), we write the leading interactions as
\begin{align}
  \Lag_{cc}\supset &\, \frac{C_X}{\Lambda_{cc}^6}X^+(\bar u\Gamma d){\cal G}_{cc} \nonumber\\
  &\, +\frac{C_{eN}}{\Lambda_{cc}^8}(\bar u\Gamma d)(\bar e\Gamma_N N){\cal G}_{cc}+{\rm h.c.},
  \label{eq:2cmain}
\end{align}
where the first operator defines the model-independent two-body recoil benchmark and we do not identify $X^+$ with a pointlike MeV-scale scalar, in view of the no-go result above. Here $\Gamma$ and $\Gamma_N$ denote Lorentz structures, and $\mathcal G_{cc}$ is a compact-$cc$ projector in the low-energy hadronic EFT. One convenient representation of its leading doubly charmed matrix elements is
\begin{equation}
  {\cal G}_{cc}=:{\cal D}_{cc,\mu}^\dagger{\cal D}_{cc}^{\mu}:,
  \qquad
  {\cal D}_{cc}^{i\mu}=\epsilon^{ijk}c^{jT}C\gamma^\mu c^k,
  \label{eq:gateMain}
\end{equation}
where $i,j,k$ are color indices, normal ordering removes vacuum contractions, $C$ is the charge-conjugation matrix, and $T$ denotes transposition. Because this projector has vanishing leading matrix elements in pions, nucleons, nuclei, and singly charmed mesons, the zero- and one-charm constraints discussed above do not apply at leading order. Residual lower-charm overlap is encoded by the leakage ratios $\epsilon_0,\epsilon_1$ defined below. The decays $\Xipp\to\Xip X^+$ and $\Xipp\to\Xip e^+N$ therefore become direct probes of a two-charm-selective interaction.

It is useful to define the following hierarchy parameters:
\begin{align}
  g_0 &\, =g_{\rm ord}, \quad g_1=g_D, \quad g_2 = g_\Xi, \nonumber\\
  \epsilon_i &\, = \left\vert\frac{g_i}{g_2}\right\vert,\quad (i=0,1),
  \label{eq:leakageParamsMain}
\end{align}
where $g_0$ is the zero-charm coupling probed by pions, nucleons, and nuclei; $g_1$ is the coupling probed by singly charmed transitions such as $D^+\to D^0X^+$; and $g_2$ is the coupling tested by $\Xipp\to\Xip X^+$ or $\Xipp\to\Xip e^+N$ at LHCb.

For LHCb to be more sensitive than the lower-charm probes, the leakage ratios must satisfy
\begin{align}
  \epsilon_0<\epsilon_0^{\rm crit}\equiv
  \frac{g_{\rm ord}^{\rm lim}}{g_\Xi^{\rm reach}},
  \qquad
  \epsilon_1<\epsilon_1^{\rm crit}\equiv
  \frac{g_D^{\rm lim}}{g_\Xi^{\rm reach}}.
  \label{eq:minimalDominance}
\end{align}
Here $g_\Xi^{\rm reach}$ is the projected LHCb coupling sensitivity shown in Fig.\,\ref{fig:twobody}, and $g_{\rm ord}^{\rm lim}$ is the profile-likelihood limit shown in Fig.\,\ref{fig:ordinaryMain}. We estimate the singly charmed bound $g_D^{\rm lim}$ by translating a branching-fraction limit into a two-body soft-recoil coupling:
\begin{equation}
  g_D^{\rm lim}(m_X)=
  \left(\frac{2\pi\Gamma_{D^+}\Br_D^{\rm lim}}
  {\sqrt{Q_D^2-m_X^2}}\right)^{1/2},
  \label{eq:gDlim}
\end{equation}
with $Q_D=M_{D^+}-M_{D^0}$. We adopt $\Br_D^{\rm lim}=10^{-4}$ as a proxy, based on the current BESIII limit on $D^+\to D^0e^+\nu_e$~\cite{BESIII:2017DplusD0enu}.

\begin{figure}[t]
  \centering
  \includegraphics[width=0.95\linewidth]{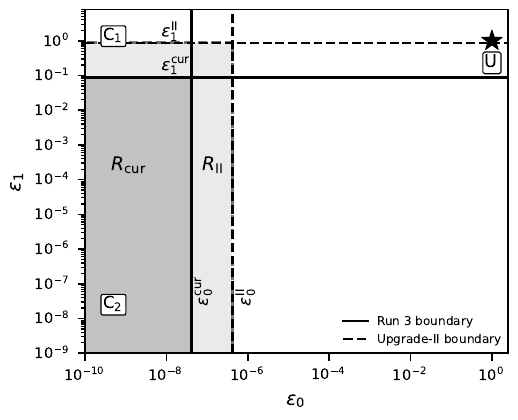}
  \caption{Dominance plane at $m_X=1\,\MeV$. The shaded rectangles $R_{\rm cur}$ and $R_{\rm II}$ indicate where LHCb provides the leading sensitivity for the product-ratio benchmarks $R_{\cal P}^{95}=10^{-3}$ and $10^{-5}$, respectively. The black star marks a universal interaction, $\mathrm U$, while $\mathrm C_1$ and $\mathrm C_2$ denote the singly charmed and two-charm-selective cases defined in Eq.\,\eqref{eq:symbolsMain}.}
  \label{fig:minimalDominance}
\end{figure}

For $m_X=1\,{\rm MeV}$ and the convention $\rho_B=1$, the combined zero-charm likelihood gives $\epsilon_0^{\rm crit}\simeq4\times10^{-8}$ for the Run-3 benchmark $R_{\cal P}^{95}=10^{-3}$ and $\epsilon_0^{\rm crit}\simeq4\times10^{-7}$ for the Upgrade-II benchmark $R_{\cal P}^{95}=10^{-5}$. The corresponding singly charmed conditions are much weaker, $\epsilon_1^{\rm crit}\simeq0.09$ and $0.9$, respectively. Fig.\,\ref{fig:minimalDominance} shows the resulting dominance plane for $m_X=1\,\MeV$. The solid and dashed shaded regions correspond to the Run-3 and Upgrade-II benchmarks, respectively. The universal ($U$), singly charmed ($C_1$), and two-charm-selective ($C_2$) cases are represented schematically by
\begin{align}
  {\rm U}:&\ (\epsilon_0,\epsilon_1)\simeq(1,1),
  \nonumber\\[-0.2em]
  {\rm C}_1:&\ \epsilon_0\ll1,\ \epsilon_1\sim1,
  \qquad
  {\rm C}_2:\ \epsilon_0\ll1,\ \epsilon_1\ll1,
  \label{eq:symbolsMain}
\end{align}
which are indicated by framed symbols in Fig.\,\ref{fig:minimalDominance}.

\section{UV interpretation and naturalness}
The $C_2$ hierarchy in Fig.\,\ref{fig:minimalDominance} can be interpreted as a selection-rule limit: when the zero- and one-charm leakage ratios vanish, the two-charm transition $\Xipp\to\Xip+{\cal P}$ remains allowed. This structure can be realized by a doubled heavy-mediator EFT with two copies $\alpha=1,2$ of charged scalars $H_\alpha^+$ and heavy vectors $V_{\alpha\mu}$.  The scalars $H_\alpha^+$, with masses $M_{H_\alpha}$, couple through $y_{q\alpha}H_\alpha^+\bar u_Rd_R$ and $y_{N\alpha}H_\alpha^+\bar N e_R$, while $V_{\alpha\mu}$ couples to the compact-$cc$ current.

An exchange symmetry with $y_{q1}=y_{q2}$, $y_{N1}=-y_{N2}$, and $M_{H_1}=M_{H_2}$ cancels the coefficient $C^{(0)}_{eN}$ of the ordinary zero-charm semileptonic operator $(\bar u_Rd_R)(\bar e_RN)$,
\begin{align}
C^{(0)}_{eN}\propto\sum_{\alpha=1}^{2}\frac{y_{q\alpha}y_{N\alpha}^{*}}{M_{H_\alpha}^{2}} = 0,
\end{align}
analogously to the Glashow--Iliopoulos--Maiani cancellation\,\cite{Glashow:1970gm}. By contrast, exchange-odd insertions in the $V_{\alpha\mu}J_{cc}$ sector generate $C^{(2c)}_{eN}$, the coefficient of the two-charm operator $(\bar u\Gamma d)(\bar e\Gamma_NN){\cal G}_{cc}$ in Eq.\,\eqref{eq:2cmain}. The explicit spurion parametrization, matching Lagrangian, and radiative-stability condition are given in Appendix\,\ref{app:uv}.

\section{Summary}
The observation of $\Xip$ establishes the doubly charmed isodoublet as an ultra-low-$Q$ laboratory for charged-current searches. We have shown that a minimal MeV-scale charged scalar is overconstrained by electroweak precision data and by pion, neutron, and nuclear beta-decay observables. A two-charm-selective interaction suppresses lower-charm transitions, allowing $\Xipp\to\Xip X^+$ and $\Xipp\to\Xip e^+N$ to become leading direct probes of the corresponding charged-current portal. The experimental handle is a topology-based feed-down analysis of reconstructed $\Xip$ candidates: In our \textsc{GenXicc}-based study, direct reconstruction of the MeV recoil or soft positron is strongly efficiency-suppressed, whereas the feed-down template reach is $R_{\cal P}^{95}\sim10^{-3}$ with the public Run-3 yield and approaches $4\times10^{-5}$ in a conditional Upgrade-II luminosity extrapolation.

\section{Acknowledgments}
This work is supported in part by the National Natural Science Foundation of China under Grant No.~12505125 and by the CAS One Hundred Talent Program.

\appendix

\section{Fast simulation and product-ratio reach}\label{app:lhcbworkflow}
This section documents the fast-simulation inputs used for the product-ratio benchmarks in Eqs.\,\eqref{eq:fastsimR95MainX} and \eqref{eq:fastsimR95MainEN}. The product-ratio sensitivity in Eq.\,\eqref{eq:RXmain} is computed as
\begin{equation}
  R_{\cal P}^{95}=
  {\mu_{95}\over N_{++}\epsilon_{\rm rel}},
  \label{eq:appFastSimReach}
\end{equation}
where ${\cal P}=X^+$ or $e^+N$ and $N_{++}$ is the reconstructed yield in the $\Xipp$ normalization channel.  For an ideal zero-background counting experiment, $\mu_{95}=3.0$. The quoted feed-down limits instead use the value of $\mu_{95}$ obtained from a template fit to prompt, cascade, and combinatorial components. The Run-3 benchmark uses $N_{++}=8712$ normalization candidates, an assumed prompt-$\Xip$ yield of $915$, and a combinatorial-background yield of $1800$.  For Upgrade II, we scale these yields by $300/6.9$, obtaining $N_{++}=3.79\times10^5$, a prompt-yield proxy of $3.98\times10^4$, and a background-yield proxy of $7.83\times10^4$.

\paragraph*{Common truth sample and production model.}
For both the two- and three-body studies, we use a $10^8$-event \textsc{GenXicc} truth-level sample for $\Xipp$ production. After retaining events with positive weights, $79\,545\,760$ events remain; the ten largest weights account for $3.28\%$ of the total positive weight. The efficiency grids use $5.0\times10^6$ decay trials per grid point and $6.0\times10^4$ events for each template study.

\paragraph*{Two-body response model.}
For ${\cal P}=X^+$, we generate isotropic $\Xipp\to\Xip X^+$ decays on a grid in $Q_\Xi$ and $m_X$, retaining only kinematically allowed points.  The benchmark detector-response model uses the LHCb forward window $2<\eta<5$, a normalization requirement $p_T(\Xipp)>4\,\GeV$, a feed-down daughter requirement $p_T(\Xip)>3\,\GeV$, and a visible-recoil trackability proxy $p(X^+)>1.5\,\GeV$. The remaining reconstruction effects are represented by explicit scale factors: $0.40$ for the normalization channel, $0.22$ for the feed-down topology, and $0.80$ for a visible recoil after the trackability requirement.  The relative efficiencies are
\begin{equation}
\begin{aligned}
  \epsilon_{\rm rel}^{\rm fd}(X^+) &={\epsilon[\Xipp\to\Xip X^+,\;X^+{\rm\ unreconstructed}]
  \over \epsilon[\Xipp\to\Lambda_c^+K^-\pi^+\pi^+]},\\
  \epsilon_{\rm rel}^{\rm vis}(X^+)&={\epsilon[\Xipp\to\Xip X^+,\;X^+{\rm\ trackable}]
  \over \epsilon[\Xipp\to\Lambda_c^+K^-\pi^+\pi^+]} .
\end{aligned}
\end{equation}

\paragraph*{Three-body response model.}
For ${\cal P}=e^+N$, events are generated according to exact three-body phase space with a constant matrix element. The feed-down efficiency is defined analogously to the two-body case,
\begin{equation}
  \epsilon_{\rm rel}^{\rm fd}(e^+N)=
  {\epsilon[\Xipp\to\Xip e^+N,\;e^+{\rm\ not\ used}]
  \over \epsilon[\Xipp\to\Lambda_c^+K^-\pi^+\pi^+]} .
\end{equation}
The direct soft-positron efficiency is
\begin{equation}
  \epsilon_{\rm rel}^{\text{soft-}e}=
  {\epsilon[\Xipp\to\Xip e^+N,\;e^+{\rm\ reconstructed}]
  \over \epsilon[\Xipp\to\Lambda_c^+K^-\pi^+\pi^+]},
\end{equation}
with the benchmark requirements $p(e^+)>1.5\,\GeV$ for trackability and $p_T(e^+)>75\,\MeV$ as a minimal electron-candidate proxy. At the baseline point $Q_\Xi=1.77\,\MeV$, $m_N=0$, our simulation gives a median $p({e^+})=9.43\,\MeV$ and a 95th percentile $p({e^+})=84.6\,\MeV$. The corresponding Upgrade-II sample gives $9.41\,\MeV$ and $84.2\,\MeV$, respectively. These values confirm that direct soft-$e^+$ counting is substantially less sensitive than the feed-down strategy.

\paragraph*{Template-fit model.}
The feed-down fit contains three components: promptly produced reconstructed $\Xip$ baryons, cascade $\Xipp\to\Xip{\cal P}$ feed-down, and combinatorial background.  Templates are constructed in $\log(1+\chi^2_{\rm IP})$ and an apparent decay-time variable $t_{\rm app}$. Here $\chi^2_{\rm IP}$ is the impact-parameter $\chi^2$ of the reconstructed $\Xip$ candidate with respect to the primary vertex, and $t_{\rm app}$ is a decay-time proxy for the reconstructed topology.

More explicitly, let $b$ label bins in the two-dimensional topology space
\[
z=\left(\log(1+\chi^2_{\rm IP}),\,t_{\rm app}\right),
\]
and denote the unit-normalized prompt, cascade, and combinatorial templates by $P_b$, $C_b$, and $B_b$, respectively. The background-only Asimov data set is then
\begin{equation}
  n_b=N_{\rm prompt}P_b+N_{\rm bkg}B_b ,
  \label{eq:templateAsimov}
\end{equation}
while the tested expectation for a cascade yield $\mu$ is
\begin{equation}
  \nu_b(\mu,\theta_p,\theta_b)=
  \theta_p P_b+\mu C_b+\theta_b B_b .
  \label{eq:templateExpectation}
\end{equation}
The nuisance normalizations $\theta_p$ and $\theta_b$ are profiled independently at each $\mu$ with the binned Poisson likelihood
\begin{equation}
  -\log{\cal L}=\sum_b\left[\nu_b-n_b\log\nu_b\right]+{\rm const.}
  \label{eq:templateLikelihood}
\end{equation}
The expected upper limit $\mu_{95}$ is defined by the one-sided profile-likelihood condition
\begin{equation}
  2\left[-\log{\cal L}_{\rm prof}(\mu_{95})
  +\log{\cal L}_{\rm prof}(0)\right]=2.71 ,
  \label{eq:templateMu95}
\end{equation}
and the feed-down product-ratio reach is $R_{\cal P}^{95}=\mu_{95}/(N_{++}\epsilon_{\rm rel}^{\rm fd})$.  The recoil object is not used in this template fit because the discrimination is provided by the reconstructed-$\Xip$ topology.  In the benchmark model, the additional upstream $\Xipp$ flight shifts the cascade component toward larger $\log(1+\chi^2_{\rm IP})$ and a longer apparent decay-time proxy than for prompt $\Xip$ production.

\begin{figure*}[t]
  \centering
  \includegraphics[width=0.95\linewidth]{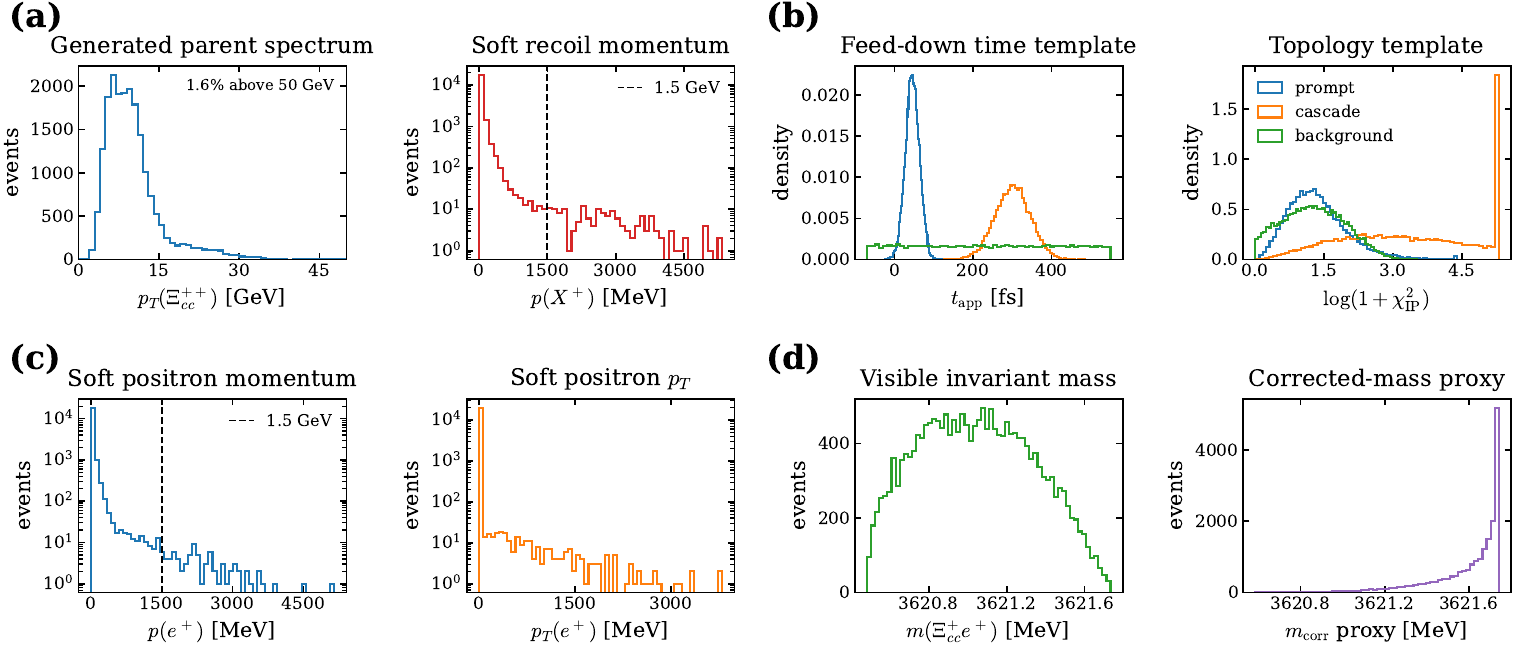}
  \caption{Representative fast-simulation diagnostics. Panels (a) and (b) show the two-body $X^+$ recoil study; panels (c) and (d) show the three-body $e^+N$ study.}
  \label{fig:fastsimDiagnostics}
\end{figure*}

Fig.\,\ref{fig:fastsimDiagnostics} collects representative Run-3 simulation diagnostics. Panel (a) shows that a MeV-scale recoil remains far below the nominal charged-track momentum scale even after boosting from the $\Xipp$ rest frame. Panel (b) displays the prompt, cascade, and combinatorial templates used in the three-component topology fit. Panel (c) shows the corresponding soft-positron kinematics for $\Xipp\to\Xip e^+N$, and panel (d) illustrates the broad visible-mass distribution together with an endpoint-like corrected-mass proxy.

Table\,\ref{tab:fastsimCard} summarizes the LHCb sensitivity used in the letter. The columns $\epsilon_{\rm rel}^{\rm fd}$ and $R_{\rm temp}^{95}$ refer to the feed-down template strategy, for which the recoil object is not reconstructed and the signal is extracted from prompt--cascade topology differences. The final two columns report a deliberately conservative direct-reconstruction proxy. The table shows that $R_{\cal P}$ can be probed through feed-down templates, whereas the large $R_{\rm obj}^{95}$ values quantify why neither a trackable MeV-scale $X^+$ nor a directly reconstructed soft positron is an effective primary observable.

\begin{table*}[t]
\caption{Numerical results from the fast simulation for $Q_\Xi=1.77\,\MeV$.  The two-body line uses $m_X=1.0\,\MeV$ and the three-body line uses $m_N=0$.  The columns $\epsilon_{\rm rel}^{\rm obj}$ and $R_{\rm obj}^{95}$ denote the direct visible-object efficiency and the ideal zero-background sensitivity for a trackable $X^+$ or a soft $e^+$, respectively.}
\label{tab:fastsimCard}
\scriptsize
\begin{ruledtabular}
\begin{tabular}{lcccccc}
scenario & ${\cal P}$ & $N_{++}$ & $\epsilon_{\rm rel}^{\rm fd}$ & $R_{\rm temp}^{95}$ & $\epsilon_{\rm rel}^{\rm obj}$ & $R_{\rm obj}^{95}$ \\
\hline
{Run 3} & {$X^+$}   & {$8.71\times10^3$} & {$0.571$} & {$1.31\times10^{-3}$} & {$2.00\times10^{-3}$} & {$1.72\times10^{-1}$} \\
{Run 3} & {$e^+N$} & {$8.71\times10^3$} & {$0.571$} & {$1.35\times10^{-3}$} & {$4.26\times10^{-4}$} & {$8.09\times10^{-1}$} \\
{Upgrade II} & {$X^+$}   & {$3.79\times10^5$} & {$0.571$} & {$3.72\times10^{-5}$} & {$1.86\times10^{-3}$} & {$4.27\times10^{-3}$} \\
{Upgrade II} & {$e^+N$} & {$3.79\times10^5$} & {$0.571$} & {$3.55\times10^{-5}$} & {$4.31\times10^{-4}$} & {$1.84\times10^{-2}$} \\
\end{tabular}
\end{ruledtabular}
\normalsize
\end{table*}

\begin{figure*}[t]
  \centering
  \includegraphics[width=0.92\linewidth]{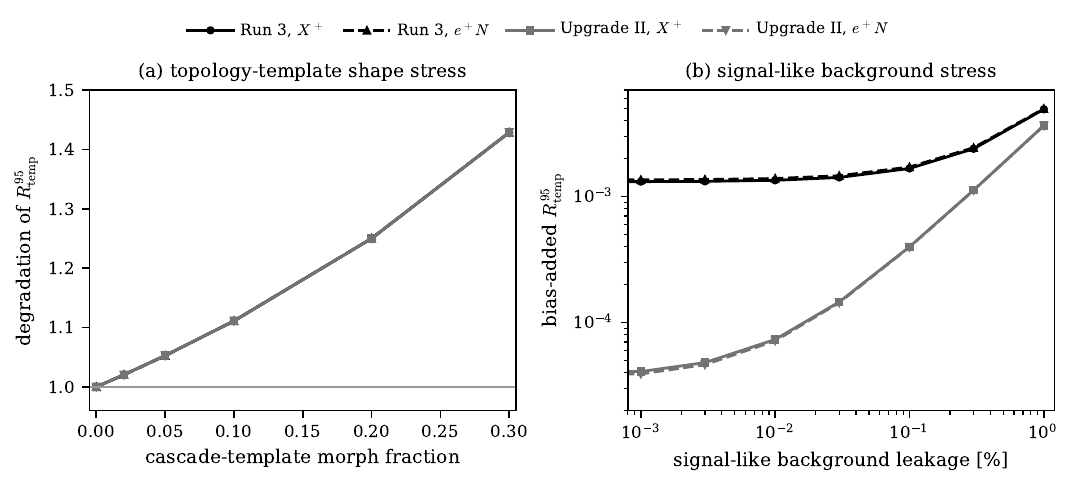}
  \caption{Representative stress tests of the feed-down template extraction.  Left: degradation of the profiled product-ratio reach when the cascade topology template is morphed toward the prompt or combinatorial-background template; for each morph fraction we show the larger of the two degradations.  Right: bias-added product-ratio reach when a signal-like component leaks into the background template.  The figure explains the entries in Table\,\ref{tab:fastsimStress}: Run-3 sensitivity is mainly limited by topology separation, while the Upgrade-II extrapolation is mainly limited by control of signal-like background leakage.}
  \label{fig:fastsimStressTests}
\end{figure*}

\paragraph*{Stress tests of topology separation.}
The feed-down observable relies on the cascade component having a topology distinguishable from prompt $\Xip$ production and from combinatorial background.  We therefore repeat the template fit under controlled deformations of the detector response and background model. In particular, we vary the impact-parameter and apparent-time responses, change the template binning, bootstrap the finite Monte Carlo templates, rescale the relative efficiency, and interpolate the cascade template toward the prompt and combinatorial shapes.  The resulting shifts are summarized in Table\,\ref{tab:fastsimStress}, while Fig.\,\ref{fig:fastsimStressTests} displays the two most relevant projections: loss of cascade--prompt shape separation and leakage of a signal-like background component.

The Run-3 projection is statistics dominated in this template framework.  Across the deformations considered here, the expected product-ratio sensitivity remains at the $R_{\cal P}^{95}\sim10^{-3}$ level for both ${\cal P}=X^+$ and $e^+N$. The leading shape effect is the cascade--prompt interpolation shown in Fig.\,\ref{fig:fastsimStressTests}(a); even a $30\%$ morph of the cascade template changes the limit by only an $\order(1)$ factor. In contrast, the luminosity-scaled Upgrade-II projection enters a systematics-limited regime.  As shown in Fig.\,\ref{fig:fastsimStressTests}(b), a signal-like background leakage at the $10^{-3}$ level would degrade the nominal $\order(10^{-5})$ reach to $R_{\cal P}^{95}\sim4\times10^{-4}$. Thus the Upgrade-II sensitivity should be interpreted as conditional on controlling signal-like feed-down backgrounds at or below the $10^{-4}$ level.

\begin{table*}[t]
\caption{Stress-test summary for the feed-down template reach at $Q_\Xi=1.77\,\MeV$.  The ``central $R_{\cal P}^{95}$'' is the benchmark value quoted in the Letter, and the ``stressed $R_{\cal P}^{95}$'' is obtained by multiplying the central value by the largest tested template-shape degradation, the largest explicit IP/time response degradation, and a $20\%$ downward shift of $\epsilon_{\rm rel}$.  The final column gives the signal-like background leakage fraction for which the fitted signal bias equals the central $\mu_{95}$.}
\label{tab:fastsimStress}
\scriptsize
\begin{ruledtabular}
\begin{tabular}{lcccc}
scenario & central $R_{\cal P}^{95}$ & stressed $R_{\cal P}^{95}$ & leakage threshold & dominant caveat \\
\hline
Run 3, $X^+$ & $1.31\times10^{-3}$ & $3.06\times10^{-3}$ & $0.362\%$ & topology separation \\
Run 3, $e^+N$ & $1.35\times10^{-3}$ & $3.23\times10^{-3}$ & $0.374\%$ & topology separation \\
Upgrade II, $X^+$ & $3.72\times10^{-5}$ & $8.87\times10^{-5}$ & $0.0103\%$ & background leakage \\
Upgrade II, $e^+N$ & $3.55\times10^{-5}$ & $8.42\times10^{-5}$ & $0.0098\%$ & background leakage \\
\end{tabular}
\end{ruledtabular}
\normalsize
\end{table*}

\section{Combined likelihood for ordinary-sector recasts}\label{app:pionlikelihood}
This section gives the individual likelihoods and their combination for the hidden-branch recast used to obtain Fig.\,\ref{fig:ordinaryMain}.

\subsection{Pion beta decay}
Let $\Br_{\pi\beta}^{\rm exp}$ denote the measured branching fraction for $\pi^+\to\pi^0e^+\nu(\gamma)$ and $\Br_{\pi\beta}^{\rm SM}$ the SM prediction. The PIBETA measurement gives~\cite{Pocanic:2003pf}
\begin{equation}
\begin{aligned}
  \Br_{\pi\beta}^{\rm exp}=\big[&\, 1.036\pm0.004_{\rm stat}\pm0.004_{\rm syst}\\
  & \pm 0.003_{\pi e2}\big]\times10^{-8},
\end{aligned}
\end{equation}
while the SM prediction and radiative-correction systematics are discussed in Ref.\,\cite{Czarnecki:2019mwq}.  If an exotic two-body mode $\pi^+\to\pi^0\phi^+$ contributes to the same selected event class with relative acceptance $\epsilon_{\pi\phi}$, a one-bin likelihood is
\begin{equation}
\begin{aligned}
-2\ln {\cal L}_\pi ={}& {\Delta_\pi^2\over \sigma_{\pi,{\rm exp}}^2 + \sigma_{\pi,{\rm unc}}^2} + \vartheta_V^2+\vartheta_R^2,\\ 
\Delta_\pi={}& \Br_{\pi\beta}^{\rm exp} - \Br_{\pi\beta}^{\rm SM}(\vartheta_V,\vartheta_R) -\epsilon_{\pi\phi}\Br_{\pi\phi} .
\end{aligned}
\end{equation}
Here $\Br_{\pi\phi}=\Br(\pi^+\to\pi^0\phi^+)$, $\vartheta_V$ is a nuisance parameter for the CKM normalization, and $\vartheta_R$ is a nuisance parameter for the short-distance radiative input.  A convenient linearized form is
\begin{equation}
 \Br_{\pi\beta}^{\rm SM}(\vartheta_V,\vartheta_R)
 =\Br_{\pi\beta,0}^{\rm SM}
 \left(1+2\sigma_V\vartheta_V+\sigma_R\vartheta_R\right),
 \label{eq:pionNuisance}
\end{equation}
where $\sigma_V$ and $\sigma_R$ are the fractional $1\sigma$ uncertainties assigned to $|V_{ud}|$ and the radiative correction, respectively.  For a scalar coupling normalized as in Eq.\,\eqref{eq:ordinaryWidth} and $m_\phi<m_{\pi^+}-m_{\pi^0}$,
\begin{equation}
\begin{aligned}
  \Br_{\pi\phi}(g_{\rm ord},m_\phi) &\, = {\tau_{\pi^+}C_\pi |g_{\rm ord}|^2\over2\pi} p_{\pi},\\
  p_{\pi} &\, = \sqrt{(m_{\pi^+}-m_{\pi^0})^2-m_\phi^2}.
\end{aligned}
 \label{eq:pionBphiLikelihood}
\end{equation}
The one-sided 95\% bound is obtained by profiling the nuisance parameters and solving
\begin{equation}
  \Delta\chi^2(g_{\rm ord})=
  -2\ln {\cal L}_\pi(g_{\rm ord})+2\ln {\cal L}_{\pi,{\rm max}}=2.71.
\end{equation}

\subsection{Neutron-decay likelihood}
For the neutron likelihood, the observable is the total decay rate.  The measured lifetime and beta-decay parameters are taken from Ref.~\cite{PDG:2024}, while the neutron--proton mass difference is also tabulated by CODATA/NIST\,\cite{NIST:CODATA2022nmp}. The likelihood is given by
\begin{equation}
\begin{aligned}
 -2\ln {\cal L}_n={}&
 {\Delta_n^2\over
 \sigma_{\Gamma_n,{\rm exp}}^2+\sigma_{\Gamma_n,{\rm unc}}^2}
 +\vartheta_V^2+\vartheta_R^2+\vartheta_A^2,\\
 \Delta_n={}&\Gamma_n^{\rm exp}
 -\Gamma_n^{\rm SM}(\vartheta_V,\vartheta_R,\vartheta_A)
 -\Gamma_{n\phi}(g,m_\phi) .
\end{aligned}
\end{equation}
Here $\vartheta_A$ denotes the neutron-specific nuisance associated with the axial-vector coupling $g_A$ and beta-correlation inputs. For the SM prediction of the neutron decay rate, we use
\begin{equation}
 \Gamma_n^{\rm SM}\propto
 G_F^2 |V_{ud}|^2 (1+3g_A^2)
 f_n(1+\Delta_R^V+\delta_R^n),
 \label{eq:neutronSMRate}
\end{equation}
Thus $\vartheta_V$ and $\vartheta_R$ are shared with the pion and nuclear likelihoods. The exotic two-body contribution is
\begin{equation}
\begin{aligned}
 \Gamma_{n\phi}(g_{\rm ord},m_\phi) &\, = {C_n|g_{\rm ord}|^2\over2\pi} p_n,\\
 p_n &\, = \sqrt{(m_n-m_p)^2-m_\phi^2} .
\end{aligned}
 \label{eq:neutronPhiRate}
\end{equation}

\subsection{Superallowed nuclear beta-like transitions}
For the nuclear likelihood, we use the corrected ${\cal F}t$ formalism for superallowed $0^+\to0^+$ transitions.  The required $Q_{EC}$ values, half-lives, branching fractions, and nuclear and radiative corrections are tabulated in Ref.~\cite{Hardy:2020qwl}.  For an isotope $i$,
\begin{equation}
  {\cal F}t_i=f_i t_i(1+\delta_{R,i}')(1+\delta_{NS,i}-\delta_{C,i}) ,
 \label{eq:FtDefinitionLikelihood}
\end{equation}
and the SM prediction is
\begin{equation}
\begin{aligned}
  {\cal F}t_{\rm SM}(\vartheta_V,\vartheta_R) &\, = {K\over 2G_F^2 |V_{ud}|^2(1+\Delta_R^V)} \\
  &\, \simeq {\cal F}t_0 \left(1-2\sigma_V\vartheta_V-\sigma_R\vartheta_R\right).
\end{aligned}
\end{equation}

The effect of a hidden branch on a published superallowed branching fraction depends on how the parent activity and the known beta branches were normalized.  We therefore introduce a channel response coefficient $\alpha_i$.  If the exotic branch is omitted from a normalization that constrains the known beta branches to sum to unity, then $\alpha_i=1$ and
\begin{equation}
 {\cal F}t_i^{\rm obs}={\cal F}t_0\,[1-\Br_{i\phi}]+{\cal O}(\Br_{i\phi}^2).
 \label{eq:FtHiddenResponseBrown}
\end{equation}
If the branching fraction is instead measured absolutely with respect to the total number of parent decays, the added width is already included and $\alpha_i\simeq0$.  A compact interpolation is given by
\begin{equation}
\begin{aligned}
 h_i(g_{\rm ord},m_\phi)&\,\equiv1-\alpha_i \Br_{i\phi}(g_{\rm ord},m_\phi),\\
 \Br_{i\phi}&\,={\tau_iC_i|g_{\rm ord}|^2\over2\pi} \sqrt{Q_i^2-m_\phi^2},
\end{aligned}
\end{equation}
for $m_\phi<Q_i$, with $\tau_i=t_{1/2,i}/\ln2$.  The sign in Eq.\,\eqref{eq:FtHiddenResponseBrown} follows directly from $t_i=t_{1/2,i}(1+P_{EC,i})/R_i$: an unobserved additional width shortens $t_{1/2,i}$ and, when omitted from the branching normalization, lowers the inferred partial half-life.

We perform a correlated, table-level recast of the 15 transitions entering the Hardy--Towner average. Let $\boldsymbol y$ denote the corrected ${\cal F}t$ values in their Table XVI and $\sigma_{i,{\rm tab}}$ the quoted transition-specific uncertainties. The covariance matrix is
\begin{equation}
 V_{ij}=\delta_{ij}\sigma_{i,{\rm tab}}^2+r_i r_j+n_i n_j,
 \label{eq:HardyCovBrown}
\end{equation}
where
\begin{equation}
 r_i={\cal F}t_i{\sigma_{\delta_R',i}\over100},
 \qquad
 n_i={\cal F}t_i{\sigma_{\delta_{NS},i}^{\rm sys}\over100}.
 \label{eq:HardyCorrVectorsBrown}
\end{equation}
The $\delta_R'$ uncertainties arise from the $Z^2\alpha^3$ term in Table V of Ref.\,\cite{Towner:2007np} and are treated as a single fully correlated nuisance, following Ref.~\cite{Hardy:2020qwl}.  The vector $\boldsymbol n$ uses the second (systematic) uncertainty quoted in Table XI of Ref.~\cite{Hardy:2020qwl}. Note that the diagonal Table-XVI uncertainties already contain the propagated $Q_{EC}$, half-life, branching-fraction, $\delta_C$, and transition-specific $\delta_{NS}$ contributions, so adding them again would double count those errors.

For fixed $(g_{\rm ord},m_\phi)$, the common conserved-vector-current (CVC) value ${\cal F}t_0$ is profiled rather than fixed.  With $\boldsymbol h=(h_1,\ldots,h_{15})$,
\begin{equation}
\begin{aligned}
 \chi^2(g_{\rm ord},m_\phi;{\cal F}t_0) &\, = (\boldsymbol y-{\cal F}t_0\boldsymbol h)^T V^{-1}(\boldsymbol y-{\cal F}t_0\boldsymbol h),\\
 \widehat{{\cal F}t}_0(g_{\rm ord},m_\phi) &\, =  {\boldsymbol h^TV^{-1}\boldsymbol y\over \boldsymbol h^TV^{-1}\boldsymbol h}.
 \label{eq:superbeta}
\end{aligned}
\end{equation}
The one-sided limit is defined by $\Delta\chi^2=2.71$. For the benchmark choice $\alpha_i=C_i=1$, the fit reproduces the CVC consistency of the survey, with $\chi^2(g=0)=6.70$ for approximately 14 degrees of freedom and $\widehat{{\cal F}t}_0=3072.30\,\mathrm{s}$, and yields $g_{\rm nuc}^{95}(m_\phi=0)=2.1\times10^{-13}$.

\subsection{Combined likelihood}
The combined analysis uses the same 15-transition data vector, response function, and Hardy--Towner covariance as the standalone nuclear fit. In particular, the nuclear likelihood is not approximated by independent isotope-by-isotope Gaussian constraints.  Using common values of $V_{ud}$ and the inner radiative correction, the predicted nuclear data vector is
\begin{equation}
\begin{aligned}
 \boldsymbol y_A^{\rm pred}(g_{\rm ord},m_\phi)&\,=
 {K\,\boldsymbol h(g_{\rm ord},m_\phi)\over2G_F^2|V_{ud}|^2(1+\Delta_R^V)},
 \\
 h_i &\,= 1-\alpha_i B_{i\phi},
\end{aligned}
 \label{eq:combinedNuclearPredictionBrown}
\end{equation}
and the nuclear contribution to the joint statistic is
\begin{equation}
 \chi_A^2=
 (\boldsymbol y_A-\boldsymbol y_A^{\rm pred})^T
 V_{\rm HT}^{-1}
 (\boldsymbol y_A-\boldsymbol y_A^{\rm pred}),
\end{equation}
where $V_{\rm HT}$ is the correlated matrix in Eq.\,\eqref{eq:HardyCovBrown}.

The pion and neutron predictions use the same $V_{ud}$ and $\Delta_R^V$:
\begin{equation}
\begin{aligned}
 \Br_{\pi\beta}^{\rm pred}={}&\Br_{\pi\beta,0}^{\rm SM} \left({V_{ud}\over V_{ud,0}}\right)^2 {1+\Delta_R^V\over1+\Delta_{R,0}^V} {1\over1+{\Gamma_{\pi\phi}\tau_{\pi^+}}}, \\
 \tau_n^{\rm pred}={}& \left[ {1\over\tau_n^{\rm SM}(V_{ud},\Delta_R^V,\lambda,K_n)} + {\Gamma_{n\phi}} \right]^{-1}.
\end{aligned}
\end{equation}
The corresponding joint chi-square is
\begin{equation}
\begin{aligned}
 \chi_{\rm comb}^2={}&{\bigl(B_{\pi\beta}^{\rm exp}-B_{\pi\beta}^{\rm pred}\bigr)^2\over\sigma_\pi^2}+{\bigl(\tau_n^{\rm exp}-\tau_n^{\rm pred}\bigr)^2\over\sigma_{\tau_n}^2}\\
 & + \chi_A^2 + z_R^2 + z_K^2 + z_\lambda^2 + z_{\pi,\rm th}^2,
\end{aligned}
\end{equation}
where $z_R$ is the single shared nuisance for $\Delta_R^V$, while $z_K$, $z_\lambda$, and $z_{\pi,\rm th}$ are channel-specific.  The common CKM normalization is profiled without an external $V_{ud}$ prior, because imposing the superallowed determination of $V_{ud}$ as an additional Gaussian constraint would double count the nuclear data.  For the nuclear-only likelihood, profiling this common normalization is algebraically equivalent, at the linearized accuracy used in the numerical implementation, to profiling the CVC value $\overline{{\cal F}t}$ in Eq.\,\eqref{eq:superbeta}.  The correlated $\delta_R'$ and $\delta_{NS}$ components are already contained in $V_{\rm HT}$ and are not added a second time.

For fixed $m_\phi$, all nuisance parameters are profiled at each nonnegative value of $g_{\rm ord}$. The test statistic is
\begin{equation}
\begin{aligned}
 \Delta\chi^2(g_{\rm ord},m_\phi)=&\,
 \chi_{\rm comb}^2(g_{\rm ord},m_\phi;\widehat{\boldsymbol\vartheta}_{g_{\rm ord}})\\
 &\,-\chi_{\rm comb}^2(\widehat g_{\rm ord},m_\phi;
 \widehat{\boldsymbol\vartheta}),
\end{aligned}
 \label{eq:combinedDchi2Brown}
\end{equation}
and the one-sided 95\% upper limit solves $\Delta\chi^2=2.71$.  For $\alpha_i=C_i=1$ and $m_\phi=0$, the updated profile gives
\begin{equation}
\begin{aligned}
 g_{\rm ord}^{95} & = 1.37\times10^{-13},\quad g_{\rm nuc}^{95} = 1.96\times10^{-13},\\
 g_{\pi+n}^{95} & = 2.30\times10^{-13}.
\end{aligned}
\end{equation}
The joint best fit is $\widehat g_{\rm ord}\simeq9\times10^{-14}$, and the null point has $\Delta\chi^2(g=0)=1.35$.

\section{UV realization of the two-charm-selective portal interaction}\label{app:uv}
\subsection{Spurion EFT and technical naturalness}
We now make explicit the construction underlying the naturalness statement in the Letter. Let $\eta_0$ and $\eta_1$ denote symmetry-breaking spurions that multiply zero- and one-charm operators, while the two-charm operator is allowed in the selection-symmetric limit.  A compact low-energy parametrization is
\begin{align}
\begin{split}
{\cal L}_{\cal P}
=
&\, \eta_0 \hat g^r_0 {\cal P}^i(\bar u_R d_R)
+\eta_1 \hat g^r_1 {\cal P}^i(\bar u_R d_R){\cal G}_{1c} \\
&\, +\hat g^r_2 {\cal P}^i(\bar u_R d_R){\cal G}_{cc}
+\text{h.c.}\, .
\label{eq:spurion_lagrangian}
\end{split}
\end{align}
Here $r=X^+,\,eN$ labels the recoil and semileptonic operator structures, ${\cal P}^{X}=X^+$ and ${\cal P}^{eN}=\bar e_R N$, and repeated $r$ indices are summed. The operator ${\cal G}_{1c}$ denotes the corresponding compact one-charm projector, while ${\cal G}_{cc}$ is the compact-$cc$ projector appearing in the main text. The observable leakage ratios used in Fig.\,\ref{fig:minimalDominance} scale as
\begin{equation}
\epsilon_i
\sim
\left|
\eta_i\,\frac{\hat g_i^r}{\hat g_2^r}
\right|
\times
\left|\frac{{\cal M}_i}{{\cal M}_2}\right|,
\qquad i=0,1,
\label{eq:epsilon_eta_relation}
\end{equation}
up to the appropriate hadronic matrix-element normalizations ${\cal M}_i$ and operator-structure dependent factors.

At the hadron level, the same structure may be represented schematically as
\begin{align}
{\cal L}^{\rm had.}_{\cal P}
={}&
g_{cc}X^+\bar T_uT_d
+G_{cc}(\bar T_uT_d)(\bar e_RN)
+y_NX^+\bar N e_R
\nonumber\\
&+\eta_1\!\left[
g_{1c}X^+\bar D^+D^0
+G_{1c}(\bar D^+D^0)(\bar e_RN)
\right]\nonumber\\
&+\eta_0\!\left[
g_0X^+\bar p n+G_0(\bar p n)(\bar e_RN)
\right]+\text{h.c.}\, .
\label{eq:hadron_spurion}
\end{align}
Under a global $U(1)_{\rm sel}$ symmetry, assign charges $+1,+1$, and $-1$ to $X^+$, $N$, and $T_d$, respectively, and take the remaining fields to be neutral.  The first line of Eq.\,\eqref{eq:hadron_spurion} is then allowed, corresponding to $\Xipp\to\Xip+{\cal P}$ and, when kinematically open, $X^+\to e^+N$.  The lower-charm terms require the compensating spurions $\eta_0$ and $\eta_1$. Thus $\eta_0,\eta_1\to0$ restores the selection rule while leaving the two-charm interaction finite. Consequently, the $C_2$ hierarchy is technically natural at the hadronic-EFT level.

\subsection{Illustrative heavy-mediator realization}
We next give an illustrative heavy-mediator EFT realization of the semileptonic operator in Eq.\,\eqref{eq:2cmain}.  This construction is intended as a gauge invariant mediator EFT below its cutoff, not as a complete gauge theory above that scale. In particular, $V_{\alpha\mu}$ denotes the heavy-vector degree of freedom in the EFT rather than a standalone Proca UV completion. Such a vector may arise, for example, as a coset gauge boson of a spontaneously broken extended gauge symmetry, with its mass generated by the corresponding Higgs sector.

Before electroweak symmetry breaking, the compact two-charm current can be embedded as
\begin{equation}
J_{cc,\mu}^{iA}
=
\epsilon^{ijk}c_R^{jT}C\gamma_\mu Q_{2L}^{kA},
\qquad
Q_{2L}=(c_L,s_L)^T ,
\label{eq:Jcc}
\end{equation}
where $i,j,k$ are color indices and $A$ is an $SU(2)_L$ index. The antisymmetric tensor $\epsilon^{ijk}$ projects the two color triplets in $\mathbf 3\otimes\mathbf 3=\bar{\mathbf 3}_A\oplus\mathbf 6_S$ onto the color-antitriplet channel.

\begin{table}[t]
\caption{
One possible $SU(3)_c\times SU(2)_L\times U(1)_Y$ assignment for the
heavy-mediator realization of the semileptonic two-charm operator.
}
\label{tab:uvcharges}
\begin{ruledtabular}
\begin{tabular}{cc}
field & SM representation \\
\hline
$V_{\alpha\mu}$, $\alpha=1,2$ & $(\mathbf 3,\mathbf 2,-5/6)$ \\
$H^+_\alpha$, $\alpha=1,2$ & $(\mathbf 1,\mathbf 1,+1)$ \\
$N$ & $(\mathbf 1,\mathbf 1,0)$
\end{tabular}
\end{ruledtabular}
\end{table}

A possible set of heavy mediator quantum numbers is shown in Table\,\ref{tab:uvcharges}, where the index $\alpha=1,2$ labels two copies of the same SM representation. The charged scalars $H_\alpha^+$ couple to $\bar u_Rd_R$ and $\bar N e_R$, while $V_{\alpha\mu}$ couples to the two-charm current. An exchange-symmetric heavy-mediator EFT may be written as
\begin{align}
{\cal L}_{\rm UV}
\supset
\sum_{\alpha=1}^{2}
\bigg(
&\, y_{q\alpha}H_\alpha^+\bar u_Rd_R
+y_{N\alpha}H_\alpha^+\bar N e_R+\kappa_\alpha V_{\alpha\mu}J_{cc}^{\mu}\nonumber\\
&\, +\lambda_\alpha H_\alpha^+H_\alpha^-V_{\alpha\mu}^\dagger V_\alpha^\mu
\bigg)+\text{h.c.}\, .
\label{eq:uv_lagrangian}
\end{align}
The fields transform under a $Z_2$ exchange symmetry as
\begin{equation}
H_1\leftrightarrow H_2,\qquad
V_1\leftrightarrow V_2,\qquad
N\to -N ,
\label{eq:Z2}
\end{equation}
with $u_R,d_R,e_R,c_R$, and $Q_{2L}$ invariant.  In the symmetric limit,
\begin{align}
\begin{split}
&\, M_{H_1}=M_{H_2},\quad
M_{V_1}=M_{V_2},\quad\\
&\, y_{q1}=y_{q2},\quad
y_{N1}=-y_{N2},\quad
\lambda_1=\lambda_2 .
\label{eq:symmetric_limit}
\end{split}
\end{align}
The leading lower-charm semileptonic coefficient then cancels:
\begin{equation}
C^{(0)}_{eN}
\propto
\sum_{\alpha=1}^{2}
\frac{y_{q\alpha}y^*_{N\alpha}}{M_{H_\alpha}^{2}} = 0,
\label{eq:C0_cancel}
\end{equation}
which is the microscopic origin of the small lower-charm spurions in Eq.\,\eqref{eq:spurion_lagrangian}.

Radiative lower-charm leakage is controlled by the exchange-odd part of Eq.\,\eqref{eq:C0_cancel}. With $s_1=+1$ and $s_2=-1$, expand near the symmetric point as
\begin{align}
\begin{split}
&\, M_{H_\alpha}^2=M_H^2(1+s_\alpha\delta_H),\quad\\
&\,y_{q\alpha}=y_q(1+s_\alpha\delta_q),\quad
y_{N\alpha}=s_\alpha y_N(1+s_\alpha\delta_N).
\end{split}
\end{align}
The induced exchange-odd lower-charm coefficient is then
\begin{align}
C^{(0)}_{eN}
&\propto \frac{2y_qy_N^*}{M_H^2} (\delta_q+\delta_N^*-\delta_H) + O(\delta^2),
\end{align}
so the relevant radiative leakage parameter is
\begin{equation}
\epsilon_0^{\rm rad}
\sim
|\delta_q+\delta_N^*-\delta_H| .
\end{equation}
If the dominant exchange-odd heavy-sector spurion is the splitting in $|\kappa_\alpha|^2$, a two-step heavy-sector feedback gives the na\"{\i}ve dimensional analysis estimate
\begin{equation}
\epsilon_0^{\rm rad}
\sim
c_{\rm rad}
\frac{\lambda\,\bar\kappa^2|\delta_\kappa|}{(16\pi^2)^2}
\frac{M_V^2}{M_H^2}
\log\frac{\Lambda_{\rm UV}}{M_V}
\log\frac{\Lambda_{\rm UV}}{M_H},
\label{eq:rad_estimate}
\end{equation}
where $\bar\kappa^2\equiv{(|\kappa_1|^2+|\kappa_2|^2)}/{2}$, $\delta_\kappa\equiv{(|\kappa_1|^2-|\kappa_2|^2)}/{(2\bar\kappa^2)}$, $c_{\rm rad}=O(1)$ encodes UV-threshold details and $\Lambda_{\rm UV}$ is the matching scale at which the heavy-mediator EFT is completed and the approximate exchange-symmetric boundary conditions are imposed. The dominance of LHCb Run-3 then requires
\begin{equation}
\lambda\,\bar\kappa^2|\delta_\kappa|
\frac{M_V^2}{M_H^2}
\lesssim 10^{-3},
\label{eq:rad_condition_run3}
\end{equation}
for comparable heavy masses and $O(1)$ logarithms, while the Upgrade-II criterion is relaxed to $\sim10^{-2}$. Equivalently, the benchmark requires collective protection, loop alignment of the exchange-odd heavy-sector spurions, or a mildly small portal combination in Eq.\,\eqref{eq:rad_estimate}.

\begin{figure}[H]
  \centering
  \includegraphics[width=0.95\linewidth]{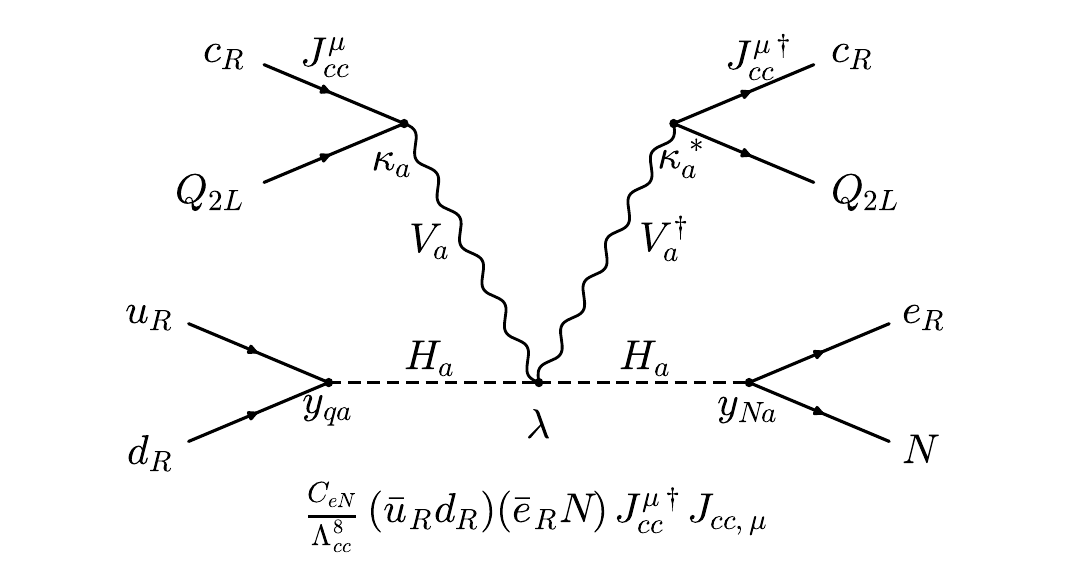}
  \caption{Tree-level matching topology for the two-charm-selective semileptonic operator.}
  \label{fig:UVmain}
\end{figure}

The desired two-charm amplitude need not vanish at tree level because it
contains the additional $V_\alpha J_{cc}$ insertions. Fig.\,\ref{fig:UVmain} shows the tree-level matching topology for one branch of the doubled mediator sector. Integrating out the heavy fields gives
\begin{equation}
\frac{C^{(2c)}_{eN}}{\Lambda_{2c}^{8}}
\sim
\sum_{\alpha=1}^{2}
\lambda_\alpha y_{q\alpha}y_{N\alpha}^{*}
\frac{|\kappa_\alpha|^{2}}
{M_{H_\alpha}^{4}M_{V_\alpha}^{4}} .
\label{eq:C2_match}
\end{equation}
In the exchange-symmetric limit for all parameters except the
exchange-odd $\kappa$ spurion, this becomes
\begin{equation}
\frac{C^{(2c)}_{eN}}{\Lambda_{2c}^{8}}
\sim
\lambda y_q y_N^*
\frac{|\kappa_1|^2-|\kappa_2|^2}
{M_H^4M_V^4}.
\label{eq:C2_odd}
\end{equation}
Thus an exchange-odd two-charm spurion can generate
$\Xipp\to\Xip e^+N$ while the leading lower-charm coefficient in
Eq.\,\eqref{eq:C0_cancel} remains absent at tree level.

Finally, relating a measured $\Xipp\to\Xip{\cal P}$ rate to the heavy
parameters requires the hadronic matrix element
\begin{equation}
\beta_\Xi^\Gamma
\equiv
\langle \Xip|(\bar u\Gamma d){\cal G}_{cc}|\Xipp\rangle ,
\label{eq:beta_xi}
\end{equation}
which should be determined from lattice QCD or estimated in heavy-diquark effective theory.

\bibliography{references}

\end{document}